\documentclass{iopart}
\usepackage[dvips]{graphicx}
\usepackage{amssymb}

\newcommand{\be}{\begin{equation}}
\newcommand{\ee}{\end{equation}}
\newcommand{\ba}{\begin{eqnarray}}
\newcommand{\ea}{\end{eqnarray}}

\newcommand{\bml}{\begin{mathletters}}
\newcommand{\eml}{\end{mathletters}}

\newcommand{\gras}[1]{\mbox{\boldmath $#1$}}
\def \lp {{\gras \lambda}}

\def \P {{\cal P}}

\def\ltsima{$\; \buildrel < \over \sim \;$}
\def\simlt{\lower.5ex\hbox{\ltsima}}
\def\gtsima{$\; \buildrel > \over \sim \;$}
\def\simgt{\lower.5ex\hbox{\gtsima}}

\begin{document}

\title[A family of filters for stochastic backgrounds...]
{A family of filters to search for frequency dependent gravitational wave stochastic backgrounds}

\author{Carlo Ungarelli
\footnote[1]{E-mail: ungarel@star.sr.bham.ac.uk} 
and Alberto Vecchio
\footnote[2]{E-mail: av@star.sr.bham.ac.uk}
}

\address{
School of Physics and Astronomy, The University of Birmingham, 
Edgbaston Birmingham B15 2TT, UK}

\begin{abstract}
We consider a three dimensional family of filters based on broken power-law spectra 
to search for gravitational wave
stochastic backgrounds in the data from Earth-based laser interferometers. We
show that such templates produce the necessary fitting factor for a wide
class of cosmological backgrounds and astrophysical foregrounds and that
the total number of filters required to search for those signals in 
the data from first generation laser interferometers operating at the design
sensitivity is fairly small.
\end{abstract}




\section{Introduction}

Gravitational wave (GW) stochastic backgrounds represent one of the classes of 
signals that could be detected 
by Earth-based laser interferometers. Stringent
upper-limits and/or detection will provide new insights into the evolution of the 
Universe at early cosmic times and high energy and, possibly, 
information about large populations of faint unresolved astrophysical 
sources at low-to-moderate redshift (see e.g.~\cite{Maggiore00} and references
therein for a recent review). The energy and spectral content of a 
stochastic background of gravitational waves are encoded in its spectrum
$\Omega(f) \equiv  (1/\rho_{\rm c})\,(d\rho(f)/d\ln f)$, where 
$\rho(f)$ is the energy density in GWs and $\rho_{\rm c}$ is 
the total energy density to close the Universe today. 
The current analysis of the data from
Earth-based laser interferometers is restricted to stochastic signals
whose spectrum is constant over the observational window, i.e. 
$\Omega (f) = $ const.
However, some theoretical models suggest that a wide class of stochastic 
backgrounds are actually described 
by a frequency dependent spectrum, peaked at some 
characteristic frequency (that depends on
the model parameters), with $\Omega (f)$ following roughly a 
"broken power-law" behaviour; this is true both for cosmological backgrounds
and astrophysical foregrounds (see for instance \cite{BMU97,FaPhy2003}). 
In the near future it is therefore important to generalise the analysis 
of the data for this class of signals.

In this paper we evaluate whether the use of
templates corresponding to  a broken power-law spectrum $\Omega(f)$
is suitable to search for stochastic backgrounds characterised by a  
fairly general frequency dependent spectrum. In particular, 
for this class of templates we investigate the fitting factor and estimate 
the related computational costs.

\section{Filter family and fitting factor}

We consider a family of  filters  corresponding to a 
class of broken power-law spectra described by 3 parameters: 
a "knee" frequency $f_s$ and two slopes $a$ and $b$. We choose the following 
functional form for the spectrum
\ba
\Omega_{\rm BPL}(f;\lp) & = & \Omega_0\,\left[A_{-}\left(\frac{f}{f_s}\right)^{a}
+ A_{+}\left(\frac{f}{f_s}\right)^{b}\right] \,,
\label{spectrum}
\ea
where $\Omega_0$ is simply a normalisation factor and does not represent 
a search parameter and the functions $A_{\pm}$ are step functions. 

The real signal will most unlikely match {\em exactly} the functional 
form~(\ref{spectrum}), and we need to explore the Fitting Factor 
(FF)~\cite{Apostolatos95} of the template family~(\ref{spectrum}).  
Lacking any solid prediction about the expected signals, we have computed the 
FF of broken power-law templates for a broad class of spectra
$\Omega(f)$. We have considered 
a Gaussian-shaped spectrum, described by the two free
parameters $f_{\rm G}$ and $\Delta f_{\rm G}$,
a Lorenzian-shaped spectrum, described by the two free parameters
$f_{\rm L}$ and $\Delta f_{\rm L}$, and a broken 
power-law spectrum modulated by an oscillatory term, whose free parameters 
are, beside $(a\,,b\,,f_s)$, the amplitude $A_{\rm osc}$ and the frequency 
$f_{\rm osc}$. The corresponding spectra read
\ba
\Omega_G(f;\lp) & = & \Omega_0\,
\exp \left[-\frac{(f-f_{\rm G})^2}{2(\Delta f_{\rm G})^2}\right]
\label{gaussian}\\
\Omega_L(f;\lp) & = & \Omega_0\,
\left[1+\frac{(f-f_{\rm L})^2}{(\Delta f_{\rm L})^2}\right]^{-1}
\label{lorenzian} \\
\Omega_{\rm osc}(f;\lp) & = & \Omega_{\rm BPL}(f;\lp)\,\times 
\left[1 + A_{\rm osc}\,\cos\left(2\pi\frac{f}{f_{\rm osc}}\right)\right]
\label{oscillation}
\ea
The above functional forms $\Omega_{\rm G}(f)$, $\Omega_{\rm L}(f)$ and
$\Omega_{\rm osc}(f)$ capture the behaviour of the spectra that characterise
a wide class of cosmological and astrophysical models (see e.g. 
\cite{BMU97,FaPhy2003}). 

\begin{figure}
\begin{center}
\mbox{
\scalebox{0.345}{\rotatebox{360}{\includegraphics{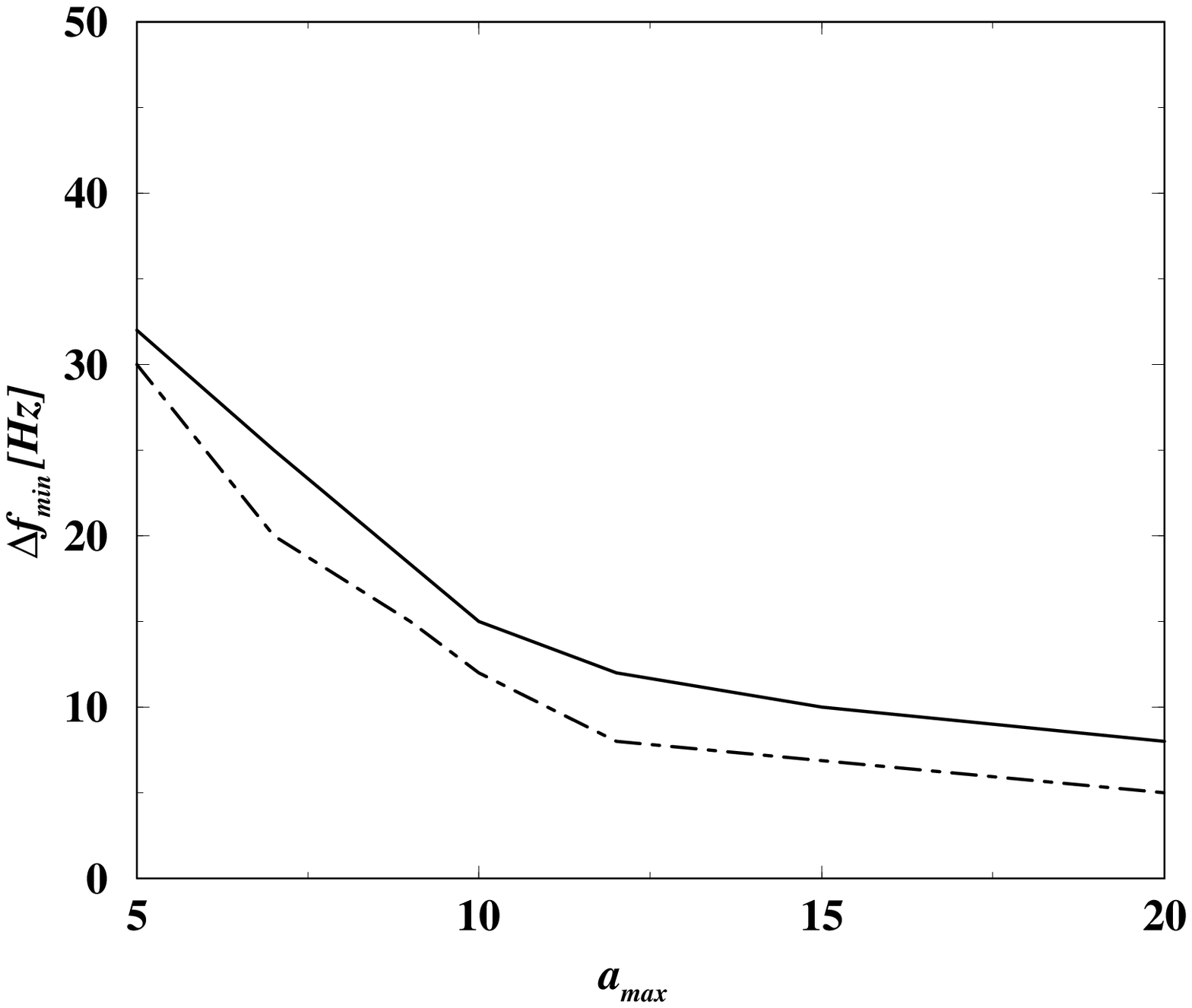}}}
\scalebox{0.345}{\rotatebox{360}{\includegraphics{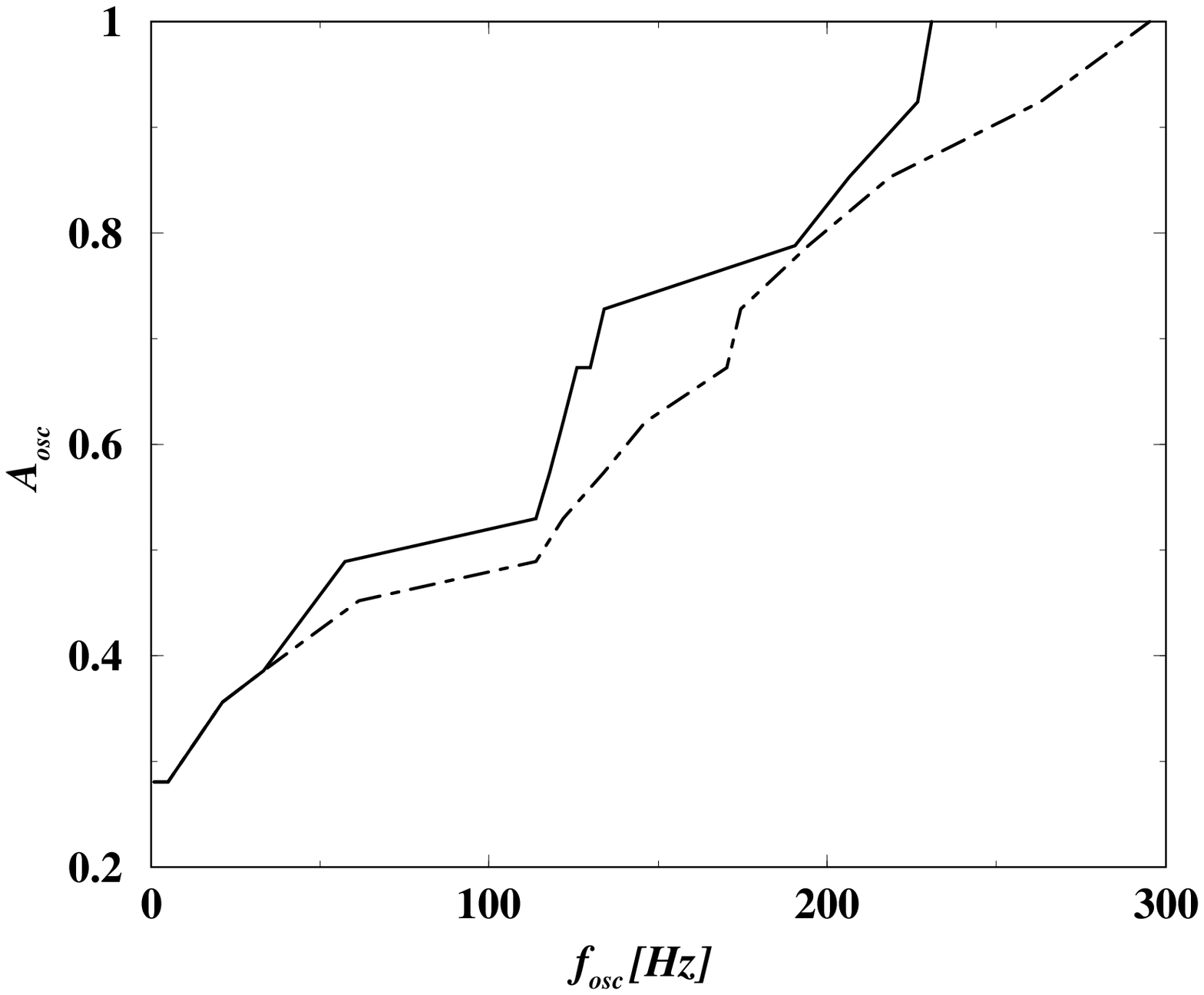}}}
}
\end{center}
\caption{
The fitting factor of ``broken power-law'' template families. 
Left: The plot shows, as a function of $ |a_{\rm max}| = |b_{\rm max}|$, 
the minimum value of $\Delta f_{\rm G}$ 
(solid line top) and $\Delta f_{\rm L}$ (dotted-line bottom) corresponding to 
$FF > 0.97$ for backgrounds characterised by a spectrum ~(\protect{\ref{gaussian}}) and~(\protect{\ref{lorenzian}}), respectively. \\
Right: The plot shows, as a function of $f_{\rm osc}$, 
the maximum amplitude $A_{\rm osc}$ for a signal~(\protect{\ref{oscillation}}) 
for which $FF > 0.97$ over a frequency band $40$Hz-$1$kHz.  The two curves correspond to a knee frequency $f_s$
in the signal corresponding to 100 Hz (top) and 1kHz (bottom), respectively. 
The slopes for both $\Omega_{\rm BPL}(f)$ and $\Omega_{\rm osc}$ are 
$a=2$ and $b=-2$.
}
\label{ff}
\end{figure}

We have computed the fitting factor of the templates $\Omega_{\rm BPL}(f)$, 
Eq.~(\ref{spectrum}), for $\Omega_{\rm G}(f)$, $\Omega_{\rm L}(f)$ and
$\Omega_{\rm osc}(f)$. The results are summarised in Figure~\ref{ff}. 
It is clear that by suitably choosing 
the parameter space for the slopes $a$ and $b$ 
one can detect essentially any spectrum characterised by a Gaussian/Lorenzian
shape. Even if the real signal shows fairly strong and rapid oscillations 
superimposed to the general power-law behaviour, the templates $\Omega_{\rm BPL}(f)$
can still have a fitting factor greater than $0.97\%$, except in extreme cases.
The main outcome of this analysis is therefore that power-law templates are
indeed suitable for searching for fairly general frequency dependent stochastic backgrounds. 

\section{Number of filters}

We can now address the computational requirements to set up a search using the
templates $\Omega_{\rm BPL}(f)$. In order to compute the total number of
filters we have extended the geometrical formalism developed in ~\cite{Owen96}
to the case of stochastic backgrounds. The basic idea is the following: 
the signal is a vector in 
the vector space of the data and the $N$-parameter family of templates traces out
an $N$-dimensional template manifold. The 
parameters themselves are coordinates on this manifold, and one can introduce a 
$N\times N$ non-flat  metric $\gamma_{jk}$ which is related to the fractional 
loss in the signal-to-noise ratio  when there is a mismatch of parameters between the 
signal and the filter. The spacing of the grid of filters is determined by the 
fractional loss due to the maximum tolerable mismatch $\mu$, which fixes the grid spacing of the filters in the parameter space $\cal P$. For a a 
hyper-rectangular mesh (which does not represent necessarily the most efficient tiling of the parameter space) the number of filters ${\cal N}$ is given by 
\be
{\cal N} = \left [ \frac{1}{2} \sqrt {N \over \mu} \right ]^{N}\,
\int_{{\cal P}} d^N{\lambda} \sqrt{\det ||\gamma_{jk}||} \,;
\label{Nf}
\ee

In particular, by adopting a detection strategy based upon  
the cross correlation statistics~\cite{AR99}, 
for a stochastic signal characterised by the spectrum
~(\ref{spectrum}) the metric $\gamma_{jk}$ is represented by a $3\times 3$ 
symmetric matrix whose components depend on the following inner product 
(involving  the optimal filter $Q$) and its 
$1^{st}$ and $2^{nd}$ derivatives with respect to the slopes $(a,b)$ and to 
the ``knee'' frequency $f_s$ (a more 
detailed derivation can be found in \cite{AV04})
\begin{equation}
\left < Q,Q\right > =\int_{f_{min}}^{f_{max}}df\,
\frac{\gamma^2(f)\Omega^2_{\rm BPL}(f; \lp)}{f^6\,S_1(f)\,S_2(f)} 
\end{equation}
where $(f_{min},f_{max})$ are the 
low-frequency and high-frequency cut-off of the interferometer band 
(chosen to be $40$Hz and $1$kHz respectively), $S_{1,2}$ 
are the noise spectral densities of the detectors and $\gamma$ is the overlap 
reduction function.

\begin{figure}
\begin{center}
\mbox{
\scalebox{0.345}{\rotatebox{360}{\includegraphics{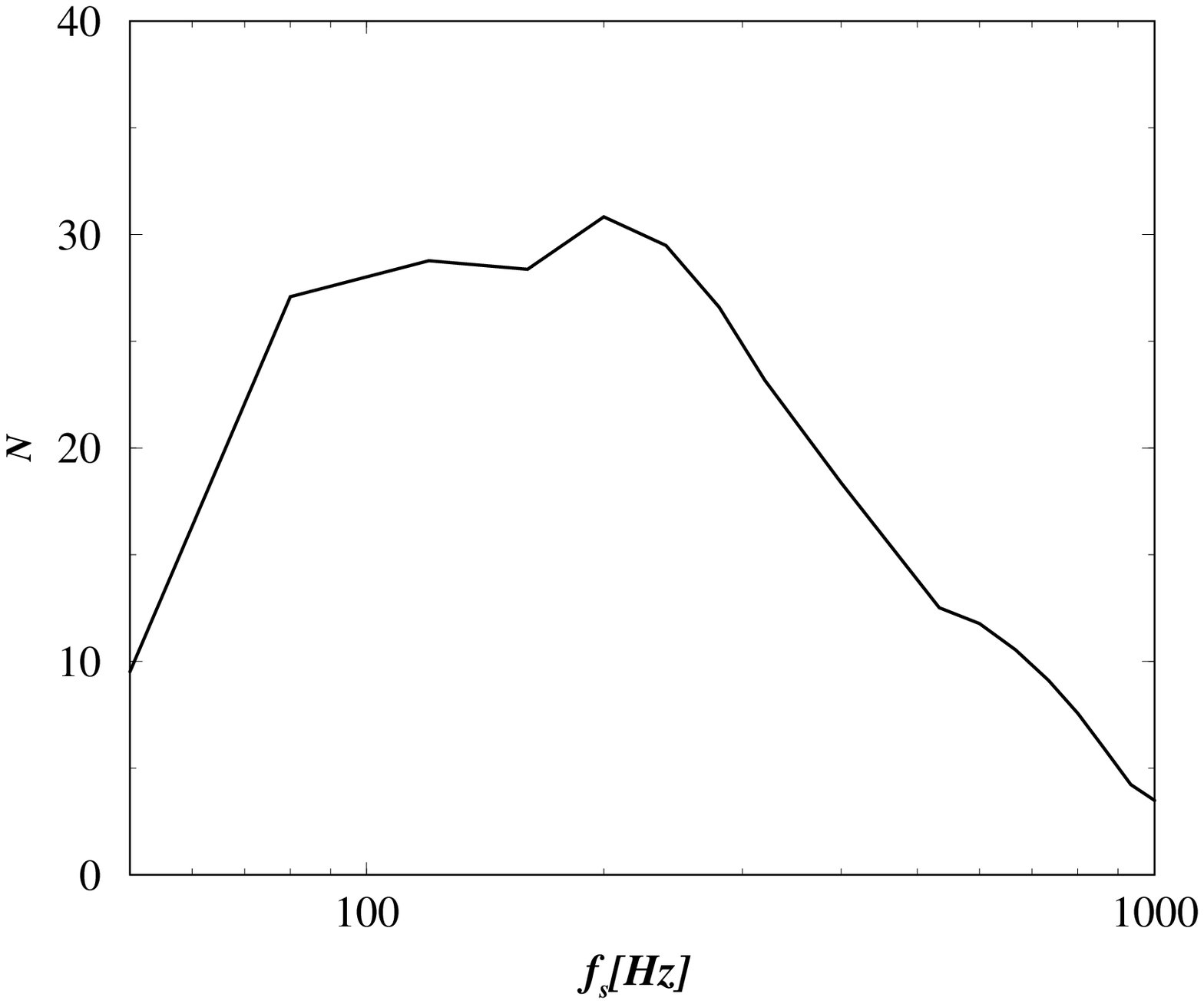}}}
\scalebox{0.345}{\rotatebox{360}{\includegraphics{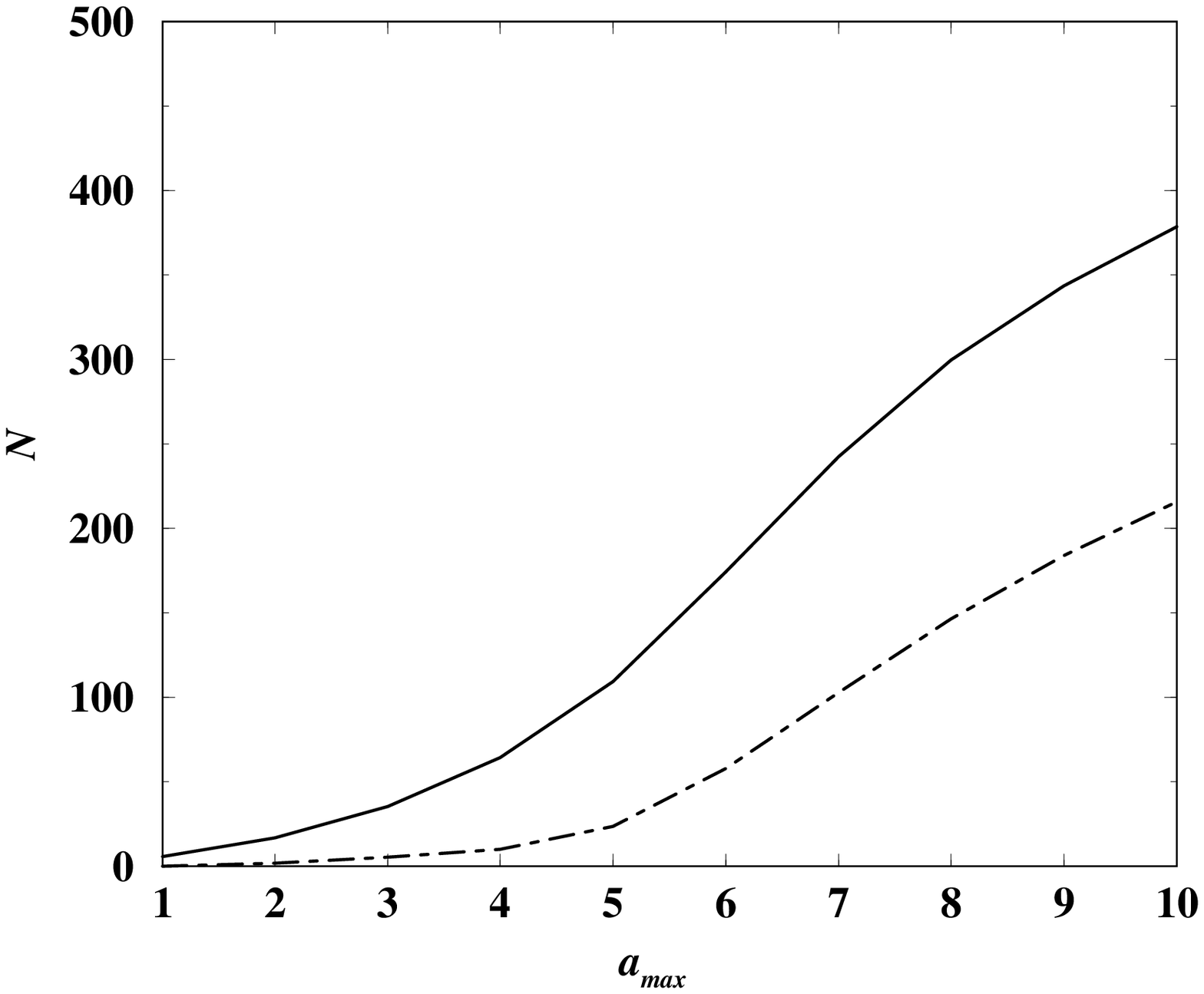}}}
}
\end{center}
\caption{
The number of filters for a search based on ``broken power-law'' templates.
Left: The plot shows the total number of filters, assuming to know 
{\it a priori} the frequency $f_s$, required to carry out the analysis
over $10^7$ sec by cross-correlating the data streams from the two
LIGO interferometers at designed sensitivity, 
as a function of $f_s$. The range of the slopes is $-10 \le a,b \le 10$.
Notice that, as expected, the largest number of filters 
corresponds to $f_s$ right at the heart of the sensitivity window. 
Right: The plot shows the total number of filters for the 
general case of a $3-$dimensional parameter search (both on the 
spectral slopes and the knee frequency).
We show the total number of filters as a function 
of the maximum value ($a_{\rm max}$) of the slope parameter $a$, 
assuming a maximum signal-to-noise ratio loss of $3\%$, 
for two different choices 
of the parameter space: (i) $a_{\rm min}=b_{\rm min}=-10$, 
$b_{\rm max}=a_{\rm max}$ (solid line), 
and (ii) $a_{\rm min}=b_{\rm max}=0$, $b_{\rm min}=-a_{\rm max}$ (dashed-dotted line). 
}\label{nfilt}
\end{figure}

We have applied this formalism to stochastic backgrounds whose 
spectrum is given by~(\ref{spectrum}) and we have computed 
${\cal N}$ for $N=3$ and $\mu = 0.03$ (a $3\%$ maximum tolerable mismatch) 
for the case of cross-correlations between
the Livingston and Hanford LIGO interferometers operating
at the designed sensitivity over $10^7$ sec of integration time. Figure~\ref{nfilt}
summarises the results. In particular, 
we have considered both the more likely  case of broken 
power law spectra with a peak in the sensitive region of the detector 
frequency band - e.g. due to a phase transition in the Early Universe 
-(dash-dot line) and the case when also 
spectra with a ``valley'' are taken into account (solid line). 
 It is clear that for a 3-dimensional 
search over the parameter space $40\,{\rm Hz} \le f_s \le 1\,{\rm kHz}$ 
and $-10 \le a,b \le 10$ - which essentially covers the relevant parameter
space (see previous section)- the total number of filters that
are required for a 3\% mismatch is $\simlt 500$. 
This represents a modest number of filters, and
the search does not pose any significant challenge as far as computational time
is concerned.

\section{Conclusions}

We have considered a family of templates characterised by a broken power-law 
spectrum to search for fairly general classes of frequency-dependent gravitational
wave stochastic backgrounds. We have shown that  templates corresponding to 
broken-power law spectra  have
the necessary fitting-factor (i.e. greater than $97\%$)  
for a wide class of models and that the 
total number of filters required to carry out a search using the data from 
first generation laser interferometer is sufficiently small not to produce
a large computational burden.

\end{document}